\documentclass[12pt]{article}

\usepackage{amsmath,amssymb,amsthm,amsfonts,mathtools}
\usepackage[margin=1in]{geometry}
\usepackage[hidelinks]{hyperref}
\usepackage{setspace}
\usepackage{enumerate}
\usepackage{natbib}
\newtheorem{theorem}{Theorem}
\newtheorem{definition}{Definition}

\newtheorem{proposition}{Proposition}

\DeclareMathOperator*{\argmax}{arg\,max}

\begin{document}

\title{Subgame Credible Nash Equilibrium}

\author{Mehmet Mars Seven\footnote{ Department of Political Economy, King's College London, London, UK. E-mail: mehmetmarsseven@gmail.com}}

\date{2 November 2025}
\maketitle

\begin{abstract}
We propose the \textit{Subgame Credible Nash Equilibrium} (SCNE), a refinement of subgame perfect Nash equilibrium (SPNE) for multi-stage games. SCNE retains the internal credibility requirement of SPNE---equilibrium behavior in every subgame---and adds an external credibility requirement across equivalent subgames: whenever a player's prescribed continuation strategy differs across equivalent subgames, her own continuation payoff must not decrease. The intuition is that credible punishments or promises should not strictly harm the punisher relative to an equivalent no-punishment subgame. The SCNE eliminates equilibria sustained by self-harming punishments or promises while preserving existence. Every multi-stage game admits an SCNE, and if each stage game has a unique Nash equilibrium, the SCNE is unique. 
\end{abstract}

\noindent \textit{JEL}: C70, D81

\noindent \textit{Keywords}: subgame perfect equilibrium, non-cooperative games, Nash equilibrium

\section{Introduction and Definition}
\label{sec:setup}

Let $G=(G^1, G^2, \ldots)$ denote a possibly infinite-horizon multi-stage game in which, at time $t$, a finite stage game $G^t=(S^t_i,u^t_i)_{i\in N}$ in normal-form is played. The finite set of players is $N$. For each $i\in N$, $u^t_i$ denotes the von Neumann–Morgenstern (vNM) utility function of player $i$ at stage $t$, and $s^t_i\in S^t_i$ denotes a (possibly mixed) action in $G^t$. When the horizon is finite, $G^T$ denotes the last stage game.

The game $G$ can equivalently be represented as $(\delta,(S_i,u_i)_{i\in N})$, where $\delta\in [0,1]$ is the common discount factor (with $\delta < 1$ if $T$ is infinite), $S_i$ is player $i$'s strategy set in the multi-stage game, $u_i$ is the discounted vNM utility of $i$, and $s=(s_i)_{i\in N}\in S=\prod_{i\in N} S_i$ is a strategy profile. 

For $k\ge1$, define a \textit{subgame} $g=(G^k,G^{k+1},\ldots,G^T)$ of $G$. Two subgames $g=(G^k,\ldots,G^T)$ and $g'=(G^{k'},\ldots,G^T)$ are called \textit{equivalent} if $k=k'$. For any strategy profile $s$, let $(s|g)$ denote its restriction to subgame $g$.\footnote{Note that $(s_i|g)$ denotes player $i$'s complete contingent plan within subgame $g$, including her behavior at the first stage of $g$, i.e., in $G^k$. For example, consider a tit-for-tat (TFT) strategy in a repeated Prisoner's Dilemma. Both players cooperate ($C$) on its path of play. Suppose that player $j\neq i$ deviates to defect ($D$) in some stage game, leading to an off-path subgame $g'$, and let $g$ denote the on-path subgame. Then, $(s_i|g')\neq (s_i|g)$ because player $i$'s prescribed first action in $g'$ is $D$, whereas in the on-path subgame $g$ it would have been $C$. This distinction clarifies that the restriction of a strategy to a subgame reflects the player's full contingent behavior from that point onward, including any adjustments in response to deviations.}

A strategy profile $s$ is a \textit{Nash equilibrium} \citep{nash1950a} if, for each player $i$,
\[
s_i \in \argmax_{s'_i\in S_i} u_i(s'_i,s_{-i}).
\]
If $(s|g)$ is a Nash equilibrium in every subgame $g$, then $s$ is a \textit{subgame perfect Nash equilibrium (SPNE)} \citep{selten1965}.

\begin{definition}
A strategy profile $s^*$ in a game $G$ is a \textit{Subgame Credible Nash Equilibrium (SCNE)} if:
\begin{enumerate}[(i)]
    \item \emph{Internal credibility:} for every subgame $g$ of $G$, $(s^*|g)$ is a Nash equilibrium; and
    \item \emph{External credibility:} for every on-path subgame $g$, every off-path subgame $g'$ equivalent to $g$, and every player $i$ who has not deviated from $s^*$ prior to $g'$,
    \[
    (s^*_i|g')\neq(s^*_i|g) \quad\Longrightarrow\quad u_i(s^*|g')\ge u_i(s^*|g).
    \]
\end{enumerate}
\end{definition}

SCNE strengthens the notion of SPNE by adding an \emph{external credibility} requirement. SPNE imposes \emph{internal credibility}: within every subgame, the profile must be a Nash equilibrium. However, SPNE does not restrict how a player's strategy may vary across equivalent subgames. In an SCNE, any threat or punishment triggered by a deviation must be credible both within the subgame where it is applied and across subgames that are strategically identical. A player's punishment strategy fails to be credible if it strictly lowers her own continuation payoff in an off-path subgame relative to her continuation payoff in an equivalent on-path subgame. Thus, SCNE refines SPNE by requiring that when a player adjusts her strategy in response to others' deviations, her continuation payoff in the resulting off-path subgame cannot be strictly lower than in the corresponding on-path subgame.\footnote{A strong SCNE is an SCNE that induces an SCNE in every subgame.}

Consider the twice-repeated game in Figure~\ref{fig:two-stage}. The profile $(C,C)$ in the first stage can be supported as an SPNE by prescribing $(E,E)$ on the equilibrium path and $(D,D)$ following any deviation. However, this profile is not an SCNE. Playing $D$ in the second stage punishes both players, lowering the punisher's own payoff relative to the on-path continuation. Because all second-stage subgames are equivalent, this off-path punishment violates external credibility. Hence, the SPNE fails to be an SCNE.

\begin{figure}[h!]
\centering
\[
\begin{array}{r|ccc}
  & C & D & E\\\hline
C & 4,4 & 0,0 & 0,5\\
D & 0,0 & 1,1 & 0,1\\
E & 5,0 & 1,0 & 3,3
\end{array}
\]
\caption{A two-stage game: an SPNE that is not Subgame Credible.}
\label{fig:two-stage}
\end{figure}

The following existence theorem is immediate. 

\begin{theorem}[Existence]
Every multi-stage game $G$ admits an SCNE.
\end{theorem}

\begin{proof}
Every $G$ has an SPNE because each stage game $G^t$ has a Nash equilibrium. Moreover, any two equivalent subgames share the same set of SPNEs. Construct an SPNE $s^*$ such that $(s^*|g)=(s^*|g')$ for all equivalent subgames $g$ and $g'$. Then $s^*$ satisfies both internal and external credibility and is therefore an SCNE.
\end{proof}

\begin{proposition}[Uniqueness]
Let $G$ be a finite-horizon multi-stage game such that each stage game $G^t$ has a unique Nash equilibrium. Then $G$ has a unique SCNE $s$ in which, for every $t$, $s^t$ is the unique Nash equilibrium of $G^t$.
\end{proposition}

\begin{proof}
If each stage game $G^t$ has a unique Nash equilibrium $s^t$, then the given profile $s$ is an SPNE, satisfying internal credibility. Moreover, since players' strategies do not differ across on-path and off-path equivalent subgames, external credibility is satisfied. Hence, $s$ is an SCNE.  In a finite horizon multi-stage game $G$, no SPNE other than $s$ can exist. Since any SCNE must also be an SPNE, $s$ is the unique SCNE of $G$.
\end{proof}

\section{Comparison with related concepts in the literature}
\label{subsec:CE_vs_RP}

This section compares the SCNE with existing refinements of subgame perfect equilibrium (SPNE) that address credibility and renegotiation concerns. The key distinction is that SCNE imposes a non-cooperative, intra-player consistency condition across equivalent subgames, whereas renegotiation-proofness and related notions impose an inter-player, Pareto consistency requirement. Neither concept generalizes the other.

A subgame perfect equilibrium $s$ is weakly renegotiation-proof (WRP) if for every subgame $g$ there exists no alternative continuation equilibrium $\hat{s}$ that Pareto-dominates $s|_g$, i.e.\ that weakly improves every player's continuation payoff and strictly improves at least one player's payoff relative to $s|_g$ \citep{farrell1989}. Strong renegotiation-proofness (SRP) strengthens WRP by requiring that no continuation of $s$ be Pareto-dominated by another WRP equilibrium. Both notions are weaker than the strong perfect equilibrium of \citet{rubinstein1980}, which extends Aumann's strong Nash equilibrium to supergames.

\citet{bernheim1989} introduce the notions of internal consistency and external consistency, based on a chain-dominance relation distinct from the external credibility condition in SCNE. In two-player games, internal consistency coincides with WRP.

SCNE rules out any strategy in which a player punishes herself and thereby reduces her own continuation payoff across equivalent subgames. In contrast, WRP eliminates continuation prescriptions that players could credibly renegotiate away through Pareto improvements among WRP-equilibrium continuations. Thus, SCNE imposes intra-player payoff consistency across equivalent subgames, while renegotiation-proofness imposes inter-player Pareto consistency. The two notions address distinct dimensions of credibility: renegotiation-proofness relies on potential communication or collective renegotiation among players, while SCNE is a fully non-cooperative refinement. Both SCNE and WRP always exist, whereas SRP may fail to exist (for example, in the Bertrand duopoly).

\subsubsection*{Example 1: SCNE does not imply WRP}

\begin{figure}[h]
\[
\begin{array}{r|c|c|}
\multicolumn{1}{r}{} & \multicolumn{1}{c}{A} & \multicolumn{1}{c}{B}\\
\cline{2-3}
A & 0,0 & 0,0\\
\cline{2-3}
B & 1,3 & 4,2\\
\cline{2-3}
\end{array}
\qquad
\begin{array}{r|c|c|}
\multicolumn{1}{r}{} & \multicolumn{1}{c}{A} & \multicolumn{1}{c}{B}\\
\cline{2-3}
A & 1,1 & 1,1\\
\cline{2-3}
B & 1,3 & 1,3\\
\cline{2-3}
\end{array}
\]
\caption{An SCNE that is not WRP.}
\label{fig:example1}
\end{figure}

Consider a two-period game with $\delta=1$. The second-period stage game (right panel of Figure~\ref{fig:example1}) yields the Row player a payoff of $1$ regardless of actions, and gives the Column player $1$ if Row plays $A$ and $3$ if Row plays $B$. Every action profile is a Nash equilibrium of the second-period game, and the WRP-efficient continuation payoff is $(1,3)$, since $(1,1)$ is Pareto-dominated.

Define the following strategy profile: the Row player plays $B$ in the first stage and, if $(B,B)$ was played, plays $B$ in the second stage; otherwise, plays $A$ in the second stage. The Column player always plays $B$ in both stages. This profile is both an SPNE and an SCNE because the Row player's second-stage threat is both internally and externally credible---her continuation payoff is $1$ both on and off the equilibrium path. However, the profile fails WRP since the off-path continuation payoff $(1,1)$ is Pareto-dominated by the alternative equilibrium payoff $(1,3)$. Hence it also fails internal consistency and strong perfection \citep{rubinstein1980}. This example also illustrates that a non-Nash stage-game outcome can be supported as an SCNE.

\subsubsection*{Example 2: WRP does not imply SCNE}
\label{subsec:wrp_not_sce}

\begin{figure}[h]
\centering
\[
\begin{array}{c|cc}
 & A & B\\\hline
A & (2,2) & (0,0)\\
B & (0,0) & (1,3)
\end{array}
\]
\caption{A WRP equilibrium that is not an SCNE.}
\label{fig:example2}
\end{figure}

Figure~\ref{fig:example2} illustrates a twice-repeated game with $\delta=1$. The stage game has two Nash equilibria, $(A,A)$ and $(B,B)$, which are Pareto efficient. The following strategy profile is an SPNE: on-path, play $(A,A)$ in both periods; off-path, play $(B,B)$ in period 2. This equilibrium is WRP and, since no WRP continuation dominates $(B,B)$, it is also SRP. However, it fails SCNE: Row who switches from the on-path $A$ to off-path $B$ across equivalent second-period subgames strictly decreases her own continuation payoff from 2 to 1, violating the external credibility requirement of SCNE.

To conclude, the concept developed in this paper, the Subgame Credible Nash Equilibrium (SCNE), extends naturally to general extensive-form games. Two subgames are said to be equivalent if their extensive forms are identical up to a relabeling of strategies, histories, and information sets. In this broader framework, an SCNE is a subgame perfect Nash equilibrium that additionally requires external credibility: whenever a player's prescribed strategy differs across equivalent subgames, her continuation payoff in an off-path ``punishment'' subgame cannot be strictly lower than her continuation payoff in the corresponding on-path subgame.

\bibliographystyle{chicago}

\end{document}